\documentclass[
 ,final            
  ]
  {aipproc}

\layoutstyle{6x9}
\newcommand{\aap}{A\&A}
\newcommand{\apj}{ApJ}
\newcommand{\mnras}{MNRAS}

\newcommand{\aj}{AJ}
\newcommand{\nat}{Nature}
\newcommand{\beq}{\begin{equation}}
\newcommand{\eeq}{\end{equation}}
\newcommand{\beqn}{\begin{eqnarray}}
\newcommand{\eeqn}{\end{eqnarray}}
\newcommand{\lppr}{\stackrel{<}{\scriptstyle \sim}}
\newcommand{\gppr}{\stackrel{>}{\scriptstyle \sim}}

\begin{document}

\title{Periodic variability and binary black hole systems in blazars}

\author{Frank M. Rieger}{
  address={Department of Mathematical Physics, University College Dublin, 
           Belfield, Dublin 4, Ireland}
}

\begin{abstract}
We consider the periodic modulation of emission from jets in blazar-type sources. 
A differential Doppler boosting origin, associated with the helical motion of a 
radiating component, is analyzed for different periodic driving sources including 
orbital motion and jet precession in a binary black hole system (BBHS). We emphasize 
that for non-ballistic helical motion classical travel time effects can lead to 
strong shortening effects, such that the observed period may be a factor $\gamma_b^2$
smaller than the underlying driving period, where $\gamma_b$ denotes the bulk Lorentz 
factor of the jet flow. The relevance of the above noted scenarios is discussed for 
the BL Lac object AO 0235+16.
\end{abstract}

\maketitle


\section{Introduction}
  Periodic variability has now been detected in the lightcurves of a significant number 
  of blazar sources, albeit often on different timescales: While some of the well-known 
  TeV blazars such as Mkn~421, Mkn~501, 3C66A and PKS 2155-304 apparently reveal evidence 
  for mid-term periodicity with timescales of several tens of days (\cite{hay98,kra01,
  lai99, oso01}) in their optical, X-ray and/or TeV lightcurves, the optical lightcurves
  from the more classical sources, e.g. BL Lac, ON~231, 3C273, 3C345, OJ~287 or AO~0235+16
  (\cite{fan98,fan01,fan02,liu95,rai01,sil88,web88}), usually suggest periods of the order
  of several years. It seems quite interesting that in many AGN the high-resolution 
  kinematic studies of their parsec-scale radio jets, particularly in several of the above 
  noted classical objects, provide strong observational evidence for the helical motion of 
  components (\cite{gom99,kel04,ste95,vic96,zen88}). This suggests that some of the observed
  periodic variabilities may arise as a result of differential Doppler boosting associated 
  with a time-dependent, periodically changing viewing angle due to motion along a helical
  jet trajectory (\cite{rie00,dep02,rie04,rie05}).   

\section{Periodic modulation of emission}
\begin{figure}
    \includegraphics[height=.3\textheight]{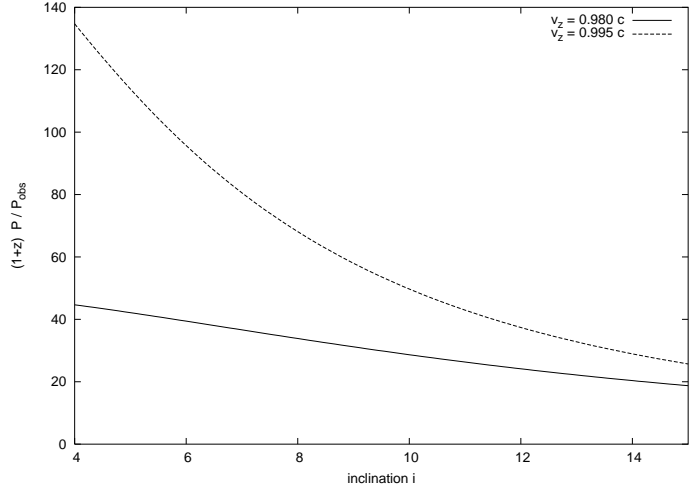}
    \caption{Ratio of real physical driving period $P$ to observed period $P_{\rm obs}$
      as a function of the inclination angle $i$ [in degrees] for two different 
      relativistic outflow velocities, corresponding to bulk Lorentz factors $\gamma_b 
      \simeq 5$ and $10$, respectively.}
      \label{fig1}
  \end{figure}

  For an emitting element moving relativistically towards a distant observer, Doppler 
  boosting effects are known to lead to a modulation of the observed flux given by
  \begin{equation}\label{modulo}
    S_{\nu}(t)=\delta(t)^{n}\,S_{\nu}'\,,
  \end{equation} where $S_{\nu}'$ is the spectral flux density measured in the comoving 
  frame, $n=3+\alpha$ for a resolved blob of plasma with spectral index $\alpha$, and
  where $\delta(t)$ is the Doppler factor depending on the actual (i.e. time-dependent) 
  angle between the velocity of the element and the direction of the observer. Obviously,
  a periodically changing viewing angle due to regular helical motion should naturally 
  lead to a periodicity in the observed lightcurves even for an intrinsically constant 
  flux.\\
  It is straightforward to show (\cite{rie04,rie05}) that in the case of non-ballistic 
  (i.e. non-radial) helical motion any underlying physical driving period $P$ will 
  appear shortened when measured by a distant observer as a consequence of classical
  travel time effects. For a relativistic outflow velocity $v_z$ along the z-axis and
  an inclination angle $i$ between the z-axis and the direction of the observer one
  finds
  \beq\label{shortening1}
     P_{\rm obs} \simeq (1+z)\,\left[1-\frac{v_z}{c}\,\cos i\right]\,P\,, 
  \eeq where $P_{\rm obs}$ denotes the observed period and $z$ is the redshift of 
  the source. It is obvious that for a sufficiently high $v_z$ and small inclination
  angles $i$, the observed periods can be much smaller than the physical driving period , 
  cf. {\bf Fig.~\ref{fig1}}.
  Moreover, for a typical blazar source with inclination angle $i \simeq 1/\gamma_b$ and 
  bulk Lorentz factor $\gamma_b \simeq (5 - 15)$, Eq.~(\ref{shortening1}) simplifies to
  \beq\label{shortening2}
     P_{\rm obs} \simeq \frac{(1+z)}{\gamma_b^2}\,P\,.
  \eeq 

\section{Possible periodic driving mechanisms}
The formation of helical jet trajectories can be related to several periodic driving
mechanisms, such as (i) the orbital motion in a binary black hole system (BBHS), (ii) 
Newtonian jet precession caused by the binary companion, or (iii) an intrinsically 
rotating jet flow. Models associated with (i) and (ii) are based on the plausibility
of close BBHSs in the centres of AGN, an assumption supported by several lines of 
arguments: Hierarchical galaxy evolution schemes, for example, suggest that BBHS
may be present in the center of elliptical galaxies as a result of mergers between 
spiral galaxies, each containing its own BH (e.g., \cite{beg80,hae02,ric98}).
Moreover, a multitude of observational evidence including the misalignment, precession  
and wiggling of extragalactic radio jets, the helical motion of radio components, and
the long- and mid-term periodicities in the optical, X-ray or TeV lightcurves has 
indeed been successfully interpreted within a binary black hole framework. Finally,
the binary concept has recently gained strong observational support by the Chandra 
discovery of two active nuclei in the merging galaxy NGC 6240 (cf., \cite{kom03} for 
a review).\\ 
Existing binary models aimed at explaining periodicity usually require very close 
systems with binary separation of the order of $10^{17}$ cm or less. Stability 
arguments against a rapid loss of orbital angular momentum via gravitational 
radiation, based on general considerations of the cosmological evolution or longterm 
periodicity, then typically suggest (Keplerian) orbital periods $P_k$ of the order 
of several years (or larger), implying observable periods $P_{\rm obs} \geq 10$ days 
(cf. Eq.~\ref{shortening2}) in the case (i) of orbital-driven (non-ballistic) helical 
motion (\cite{rie04}). In the case (ii) of Newtonian jet precession the corresponding
periods will be much larger. Newtonian precession may arise due to tidally induced 
perturbations in the accretion disk of the jet-emitting primary source caused by the
binary companion. Under favourable conditions these perturbations can lead to a 
rigid-body precession of the inner parts of the disk and thus translate into a 
precession of the jet (\cite{kat97,lar98,rom00,rom03}). 
The implied physical precessional driving period $P_p$ is usually a factor ten (or 
more) larger than the orbital period $P_k$ of the binary (cf. \cite{rie04}), 
suggesting that non-ballistic helical motion due to Newtonian jet precession is 
unlikely to be responsible for periodicity on a observed timescale of less than
one hundred days, but may well be associated with observed periods $P_{\rm obs} 
\gppr 1$ yr.\\
In general, the helical motion of components does not necessarily require a BBHS. 
An intrinsically rotating jet (case (iii)) for example, may also mimic some of 
the observational signatures, provided components are dragged with the underlying 
rotating flow. The occurrence of such an internal jet rotation (at least initially) 
appears not unlikely: The strong correlation between the disk luminosity and bulk 
kinetic power in jets and the phenomenological evidence for a jet-disk symbiosis 
(e.g., \cite{fal95,raw91}) for example, suggests that a significant amount of 
accretion energy, and hence rotational energy is channeled into the jet. 
Moreover, internal jet rotation is a natural consequence if jets are formed as
magnetized disk winds (e.g., \cite{cam92}). Information about the underlying 
rotation profile may then be used to derive possible observable periods. It can 
be shown (\cite{rie04,rie05}) for example, that for the lighthouse model of 
Camenzind \& Krockenberger~(1992) bounds on the maximum jet radius derived from 
numerical simulations translate into characteristic periods of $P_{\rm obs} \lppr 
10$ days (for massive quasars) and $P_{\rm obs}\sim 1$ day for typical BL Lac 
objects.

\section{Application to AO 0235+16}
The BL Lac object AO 0235+16 (PKS 0235+164) at redshift $z=0.94$ is well-known for
its extreme variability at almost all wavelengths. Observations of its radio structure 
at ground-based (e.g., \cite{che99,chu96}) and space (\cite{fre00}) VLBI resolutions 
have shown that AO~0235+16 is very compact on submilliarcsecond angular scales, and 
provided evidence for a very high brightness temperature in excess of $5.8 \cdot 
10^{13}$ K and high apparent superluminal motion up to $(27 \pm 6)\,c$, indicating 
very small viewing angles and large Doppler factors (cf. also \cite{fuj99}). There 
are some indications that the radio outbursts are associated with the formation of 
new VLBI components (\cite{baa84,chu96}. Based on the variation in the position angle 
of the radio jet, Zhang et al.~(\cite{zha98}) have suggested that the jet may be 
rotating and the central engine precessing (cf. also \cite{che99}). A rough 
order-of-magnitude estimate for the physical parameters in the emitting region may 
be obtained by fitting the multiwavelength SED with a homogeneous, one-zone 
synchrotron - inverse Compton model, suggesting viewing angles $\sim 2.9^{\,\circ}$, 
magnetic field strengths $\sim 3.8$ Gauss and bulk Lorentz factors $\sim 16$ for the 
low state (\cite{pad04}).\\
The analysis of the long-term variability in AO~0235+16 over a time range of $\sim 25$ 
yr has revealed evidence for a $(5.7\pm 0.5)$ yr periodicity in its radio lightcurves 
and a possible $(2.95\pm 0.15)$ yr periodicity in its optical lightcurves, e.g. 
\cite{fan02,rai01,roy00,web00}. Two scenarios have been proposed recently in order 
to account for these findings, both assuming the presence of a close BBHS:
\begin{enumerate}
\item Romero, Fan \& Nuza~(2003) have argued that AO~0235+16 may harbour a BBHS 
with the optical periodicity being related to the companion crossing the accretion disk
around the jet-emitting black hole on a non-coplanar circular orbit (hence implying an 
orbital period of $P_k \simeq 2 \times 2.95/[1+z] \sim 3$ yr) and the radio periodicity 
being related to Newtonian jet precession.
\item Ostorero, Villata \& Raiteri~(2004) on the other hand, have argued that both, the 
radio and optical periodicity (assuming the optical period to be the same as the radio!) 
may be associated with a helically bent, steadily emitting inhomogeneous jet, driven by 
the orbital motion in a close BBHS.
\end{enumerate}
If scenario~(1) is indeed realized, the fluid motion must be non-ballistic as otherwise 
the precessional period would be to short to be generated via tidally induced 
perturbations. As pointed out above, the ratio of precessional to orbital period is 
usually of the order of ten or larger, i.e. one has $P_p \gppr 30 $ yr. Provided the 
jet is not strongly inhomogeneous and the cone opening angle sufficiently small, this 
period will appear shortened when measured by a distant observer following 
Eq.~(\ref{shortening2}), thus indicating that one requires bulk flow Lorentz factors 
$\gamma_b \gppr 3.2$. On the other hand, for bulk Lorentz factors of the order $\sim 10$, 
as suggested from the studies above, the precessional driving period would be $P_p \sim 
300$ yr. The projected wavelength of the associated helical trajectory $\lambda \simeq 
P_p\,c\,/\gamma_b$ should then be of order $3$ parsec (for $P_p \simeq 30$ yr) and $9$ 
parsec (for $P_p \simeq 300$ yr), or $0.36$ mas and $1.1$ mas, respectively (assuming 
$q_0 = 0$ and $H_0=65$ km s$^{-1}$ Mpc$^{-1}$), and thus likely to be accessible for 
high-resolution VLBI observations. On the basis of scenario~(1) it may also be useful 
to search for signs of quasi-periodic variability on the timescale of several months or 
less in the high energy range. For apart from a periodic modulation due to precession, 
one may also expect the orbital motion of the binary to lead to some quasi-periodic 
modulation, at least from the initial parts of the jet where its width is still smaller 
than the separation of the binary. For a Keplerian period $P_k \simeq 3$ yr, possible
observable variability timescales range from $\sim 7$ months (for $\gamma_b \sim 3$) 
to $\sim 20$ days (for $\gamma_b \sim 10$) or perhaps even less.\\
Note that central BH masses for BL Lac objects, estimated from observations of their 
host galaxies, usually fall within a mass range $6 \cdot 10^7\,M_{\odot} \lppr (M+m) 
\lppr 10^9\,M_{\odot}$ (e.g., \cite{fal03,wu02}), a range consistent with constraints 
derived for AO~0235+16, thus suggesting a binary separation of $d \lppr 3\cdot 10^{16}$ 
cm.

The situation may however be quite different if a scenario following (2) is correct. 
At first glance such a scenario seems to be associated with the requirement that the 
observed timescale for the optical periodicity coincides with the one for the radio 
periodicity, and may thus appear less plausible if the difference suggested above is 
indeed confirmed by further observation and analysis. It is likely however, that the 
real case is much more complex: For a helically bent, steadily emitting inhomogeneous 
jet driven by the orbital motion, high energy observations probing the smallest 
scales are expected to provide the most useful tracers of the underlying Keplerian 
period. At radio energies, the corresponding jet flow will repeatedly approach the 
line-of-sight along its helical path, leading to a maximization of beaming effects 
(and thus offering a possible interpretation for the detected radio knots). Unless 
the radio jet is very inhomogeneous, the physical orbital period in the radio 
band will thus appear strongly shortened when measured by a distant observer as shown 
above, i.e. the real Keplerian period of the binary may be much larger than the 
observed radio period, an effect not considered by Ostorero et al.~(2004). For an 
observed radio period of $5.7$ yr, for example, the real Keplerian period may be 
in the range between $P_k \simeq 26$ yr (for $\gamma_b =3$) and $P \sim 300$ yr 
(for $\gamma_b =10$), implying a binary separation of $d \gppr 2 \cdot 10^{17}$ cm 
assuming the mass range given above. The observed radio lightcurves may then be 
characterized by pronounced peaks separated by $(1+z)\,P_k$, with intermediate 
peaks occurring on a timescale of $5.7$ yr.

\section{Conclusion}
There is mounting evidence that the mid- and long-term periodicity observed in
blazar-type sources is related to the presence of close BBHSs in their centres.
Here we have shown that the observed timescales of periodicity may carry valuable 
information about their physical nature.
Continuous observations in different energy ranges, a thorough periodicity 
analysis of their lightcurves and detailed theoretical modelling will allow
to shed more light on their histories and properties.


 \begin{theacknowledgments}
Dicussions with Peter Duffy and support by a Marie-Curie Individual Fellowship 
(MCIF - 2002 - 00842) are gratefully acknowledged.
 \end{theacknowledgments}


\bibliographystyle{aipprocl} 

\end{document}